\documentclass[,final]            
  {aipproc}
\layoutstyle{8x11double}

\begin{document}

\title{Interpreting black hole QPOs}

\author{Marek A. Abramowicz}{
  address={Department of Astrophysics, Chalmers University, G\"oteborg, Sweden}
}
\author{W{\l}odzimierz}{
  address={Copernicus Astronomical Center,
   ul. Bartycka 18, 00-716 Warszawa, Poland}
}
\author{Klu{\'z}niak}{
  address={Institute of Astronomy,
   Zielona G\'ora University, ul. Lubuska 2, 65-265 Zielona G\'ora, Poland}
}

\begin{abstract}
In all the microquasars with two hHz QPOs, the ratio of the
frequencies is 3:2, supporting our suggestion that a non-linear resonance
between two modes of oscillation in the accretion disk plays a role
in exciting the observed modulations of the X-ray flux.
We discuss the evidence in favor of this interpretation, and we relate
the black hole spin to the frequencies expected for various types of
resonances that may occur in nearly Keplerian disks in strong gravity.
For those microquasars where the  mass of the central X-ray source
is known, the black hole spin can be deduced from a comparison of the
observed and expected frequencies.
\end{abstract}

\maketitle


\section{Introduction}

\noindent 
Several Galactic low-mass X-ray binaries (LMXBs)
exhibit quasi-periodic variability (QPOs)
of their X-ray fluxes, with pairs of $\approx 1\,$kHz frequencies 
typical in the neutron-star sources \cite{1}.
 Klu\'zniak and Abramowicz \cite{3,3a}
suggested that these twin kHz QPOs are a manifestation of
non-linear resonance that can occur between modes of oscillation
in an accretion disk in strong field Einstein's gravity, but not in Newton's
$1/r$ potential, and pointed out that on this hypothesis the same
resonances should be present in black hole systems. Pairs
of high frequency QPOs should have been present where only single hHz QPOs
had been reported in microquasars. Such pairs of several
hHz frequencies have now been reported in four or five black-hole
systems, all in a 3:2 ratio \cite{2}, substantially strengthening the case
for resonance. 

Our originally suggested \cite{3} explanation for kHz QPOs
in neutron stars was based on these general properties of non-linear
resonance, which seemed to us to correspond to the essential features
of the observed twin frequency peaks:
\begin{enumerate}
\item The frequencies of non-linear oscillations $\nu$ depend on amplitude,
 and for this reason  they may be time dependent and may differ
from the fixed eigenvalue frequencies $\nu{(0)}$ of the system, 
$\nu (t) = \nu{(0)} + \delta \nu(t)$. 
\item Non-linear resonance may occur over a wide frequency range $\delta \nu$.
\item Both (resonant) frequencies increase or decrease ``in step'' with
 each other.
\item The eigenfrequencies of resonant modes are approximately in the ratio of
 small integers, e.g., 2:1.
\end{enumerate}
These ideas initially
received a cool reception (see author's note in \cite{3a}),
because it was not generally appreciated that frequency ratios close to 
3:2 actually occur\footnote{Indeed, our paper on this appeared in print 
only much delayed \cite{6}.} 
for kHz QPOs in neutron star
sources, and no evidence for twin peaks in black hole sources was known
at the time.

In addition to the high frequency QPOs, features in the power density spectra
can be identified at lower frequencies, and at least one frequency
$\nu_{\rm low}$ was long known to be correlated with $\nu_{\rm high}$,
one of the kHz/hHz frequencies in neutron-star/black-hole systems \cite{4}.
Quasi-periodic modulations of the flux (dwarf nova oscillations, DNOs)
were first discovered in cataclysmic variables (white dwarfs)
and these are analogous in many respects to the QPOs in LMXBs \cite{91}.
However, unlike the kHz/hHz QPOs, the highest frequency DNOs do not
come in pairs. This is consistent with the idea that the high frequency pairs
arise in accretion disks only in strong gravity.

The relativistic resonance model of black hole QPOs
is based on fundamental features of strong gravity.
Today, it
is motivated by observations
that sharply illuminate the physical nature of QPOs:       
\begin{enumerate} 

\item The correlation $\nu_{\rm low} = 0.08 \nu_{\rm high}$ between low and
 high frequency QPOs in black hole, neutron stars, and white dwarf sources 
extending over six orders of magnitude \cite{4,5,92}, proves that in general
the QPOs are a hydrodynamic phenomenon, and cannot be attributed to mere
kinematic effects, such as Doppler modulation of emission
from isolated bright spots.
($\nu_{\rm low}$ may be  the ``ninth wave'' \cite{38}.)

\item The frequencies of twin peak hHz QPOs in microquasars seem to
scale with mass \cite{2}, $\nu \sim 1/M\,$ (Figure 1). If true, this
would prove their relativistic origin.

\item In all four microquasars with twin peak hHz QPO pairs, 
$\nu_{\rm upper}/\nu_{\rm lower} = 3/2\,$ (Table 1). 

\end{enumerate}

Although suggestive of a resonance, the ratio 3/2 could also be a signature
of overtones (flute modes) \cite{36}, 
or of higher modes of an MHD instability  \cite{37} at a
`transition radius' $r_*$ in the innermost part of the disk, which
excites quasi periodic oscillations with mode frequencies 
$\nu \sim n\,\nu_{\rm K}(r_*)$ 
(in contrast with the resonance model,
neither of these two models predicts $1/M$
scaling, without making some {\it ad hoc} assumptions).
However, there are additional
properties of non-linear resonances which may help in their identification.
In a non-linear resonance
 combination frequencies, e.g., $\nu_{\rm upper} \pm \nu_{\rm lower}$,
and subharmonic frequencies may be present
\cite{93}, e.g., $\nu_{\rm lower}/2$.

 Our resonance model may also be applied to twin peak QPO sources
in neutron stars \cite{93}. In refs. \cite{12, 13}, 
and in these Proceedings \cite{14}, 
we discuss a resonance in an accretion disk or torus excited by
an external forcing by a millisecond pulsar. A similar forcing is
crucially important in a different, non-relativistic
 resonance model suggested by Titarchuk \cite{10}. 

\section{The orbital and epicyclic motions.}

Consider a black hole\footnote{We rescale mass with  
 $M = GM_0/c^2 =r_{\rm G}$, angular momentum with $a = J_0c/(M_0^2G)$.  We use
Boyer-Lindquist coordinates,  $t, r, \theta, \phi$, and rescale the
radius with  $x = r/r_{\rm G}$.} with the mass $M_0$ and angular
momentum $J_0$.  Inside thin, almost Keplerian accretion disks, matter
spirals down the central black hole along stream lines that are
located almost on the equatorial plane $\theta = \theta_0 = \pi/2$,
and that locally differ only slightly from a family of concentric
circles $r = r_0 = {\rm const}$. The small deviations, $\delta r = r - r_0$,
$\delta \theta = \theta - \theta_0$ are governed, with accuracy to
linear terms, by


\begin{equation}
\label{eq_mot_vdir}
\delta \ddot r + \omega_r^2 \,\delta r = \delta a_r ,
~~~~
\delta \ddot \theta + \omega_{\theta}^2\,\delta \theta = \delta a_{\theta} .
\end{equation}

\noindent Here, the dot denotes a time derivative. For purely Keplerian
(free) motion $\delta a_r = 0$, $\delta a_{\theta} = 0$ and the above
equations describe two uncoupled harmonic oscillators with the
eigenfrequencies $\omega_{\theta}\equiv2\pi\nu_{\theta}$,
$\omega_r\equiv2\pi\nu_r$ shown together with the Keplerian orbital
frequency, $\Omega\equiv2\pi\nu_{\rm K}$, in Figure 2 for a
non-rotating black hole, and in Figure 3 for a moderately rotating one.


\begin{table} [!ht]
\begin{tabular}{lll}
\hline
    \tablehead{1}{r}{b}{Microquasar}
  & \tablehead{1}{r}{b}{$\nu_{\rm upper}$}
  & \tablehead{1}{r}{b}{$\nu_{\rm lower}$} \\
  
\hline
~~XTE  1550-564           & ~~~~~~~~~~~276 & ~~~~~~~~~~~174 \\   

~~GRO  1655-40            & ~~~~~~~~~~~450 & ~~~~~~~~~~~300 \\ 

~~GRS  1915+105           & ~~~~~~~~~~~168 & ~~~~~~~~~~~113 \\ 

~~~~~~~H 1743-322         & ~~~~~~~~~~~240 & ~~~~~~~~~~~160 \\ 
 
\hline
\end{tabular}
\caption{The four microquasars in which two hHz QPOs are observed. 
They all have 3:2 ratio of frequencies. Source of data: \cite{2, 15}}
\label{table-1}
\end{table}


\begin{figure} [!h]
  \includegraphics[angle=-90, width=78mm]{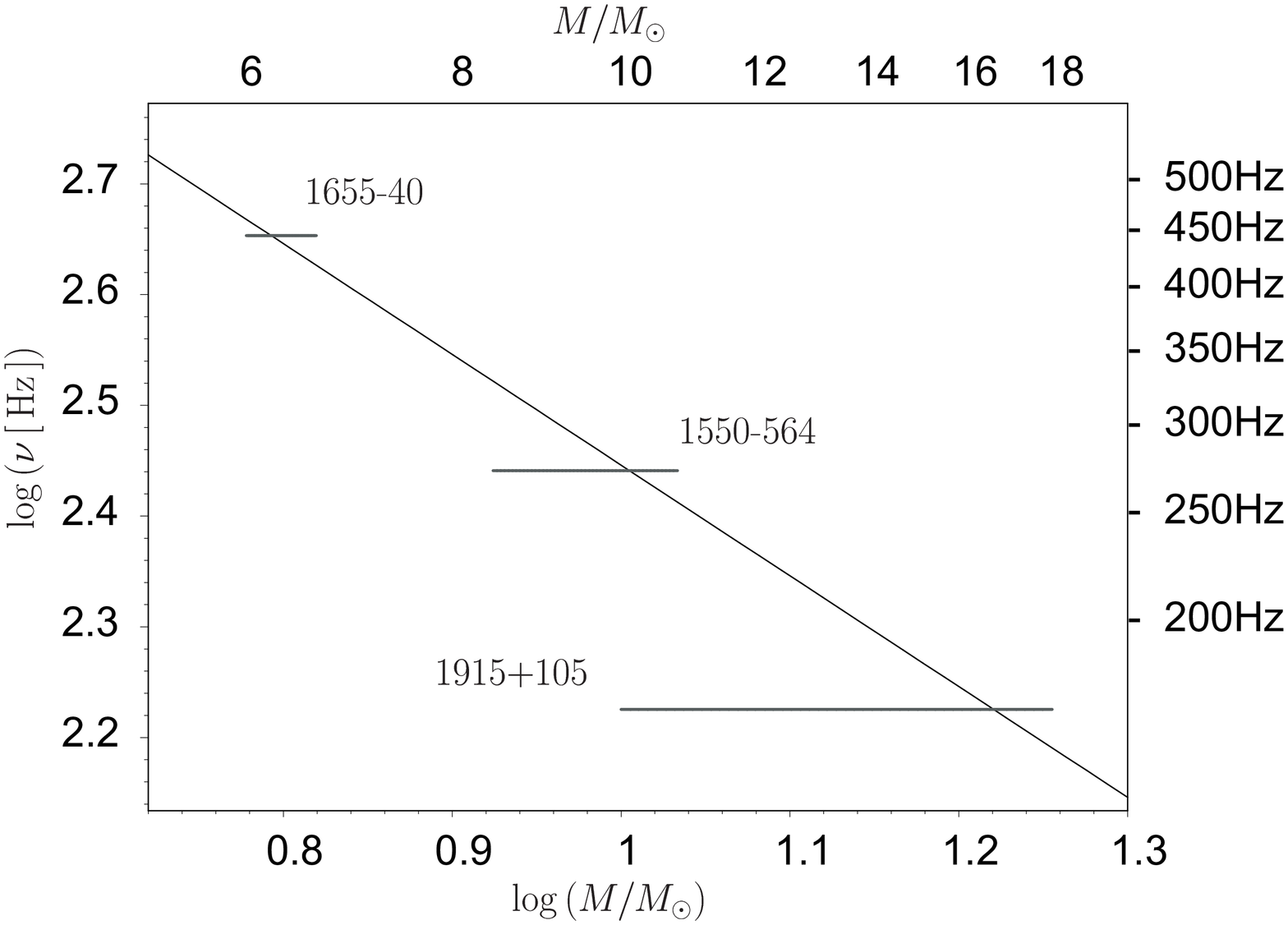}
  \caption{The $1/M$ scaling of the pairs of QPOs with the 
        3:2 frequency ratio \cite{2}. The upper frequency is shown.}
\end{figure}

\noindent In Newton's theory with the $-GM_0/r$ potential, 
$(2\pi)^{-1}GM_0/r^{3/2} = \nu_{\rm K} = \nu_r = \nu_{\theta}$, 
but in the strong gravity of a rotating black hole,
for orbits of the same sense of rotation,
 $\nu_{\rm K} >  \nu_{\theta} > \nu_r$.
The radial epicyclic frequency $\nu_r$ goes to zero at
the Innermost Stable Circular Orbit for the Keplerian (free) motion, 
and has a maximum at a particular circular orbit outside the ISCO, 
its location depends on the black hole spin 
\cite{17,16,18}.

\section{$1/M$ scaling}

\noindent
 Before the RXTE era,
Klu\'zniak, Michelson, Wagoner \cite{94} suggested that the
orbital frequency close to the marginally stable orbit may be directly observed
as a QPO, once instruments with sufficiently high time resolution are built,
and pointed out that the frequency will be inversely proportional to the
mass of the compact object. The latter statement applies to any characteristic
frequency in general relativity.

\begin{figure} [!ht]
  \includegraphics[angle=-90,
  width=78mm]{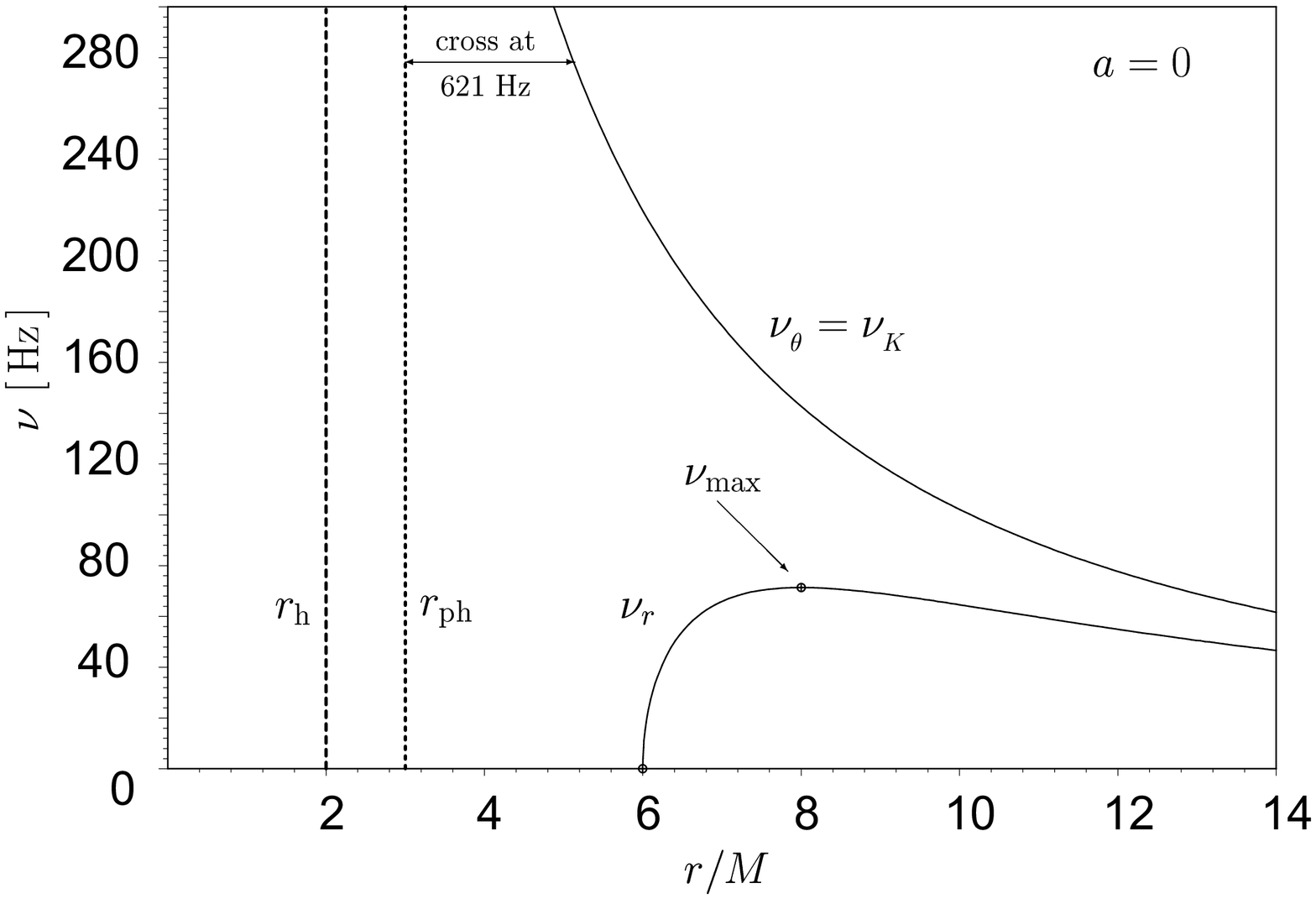}
  \caption{Orbital  $\nu_{\rm K}$, epicyclic radial
  $\nu_{\rm r}$, and vertical $\nu_{\theta}$ frequencies for a
  non-rotating black hole (spin $a=0$) with the mass 
  $M_0  =10\,M_{\odot}$. Location of the photon radius is marked by 
  $r_{\rm ph}$, and that of the horizon by $r_{\rm h}$. There are no orbits
  for $r < r_{\rm ph}$.}
\end{figure}


\begin{figure} [!ht]
  \includegraphics[angle=-90, width=78mm]{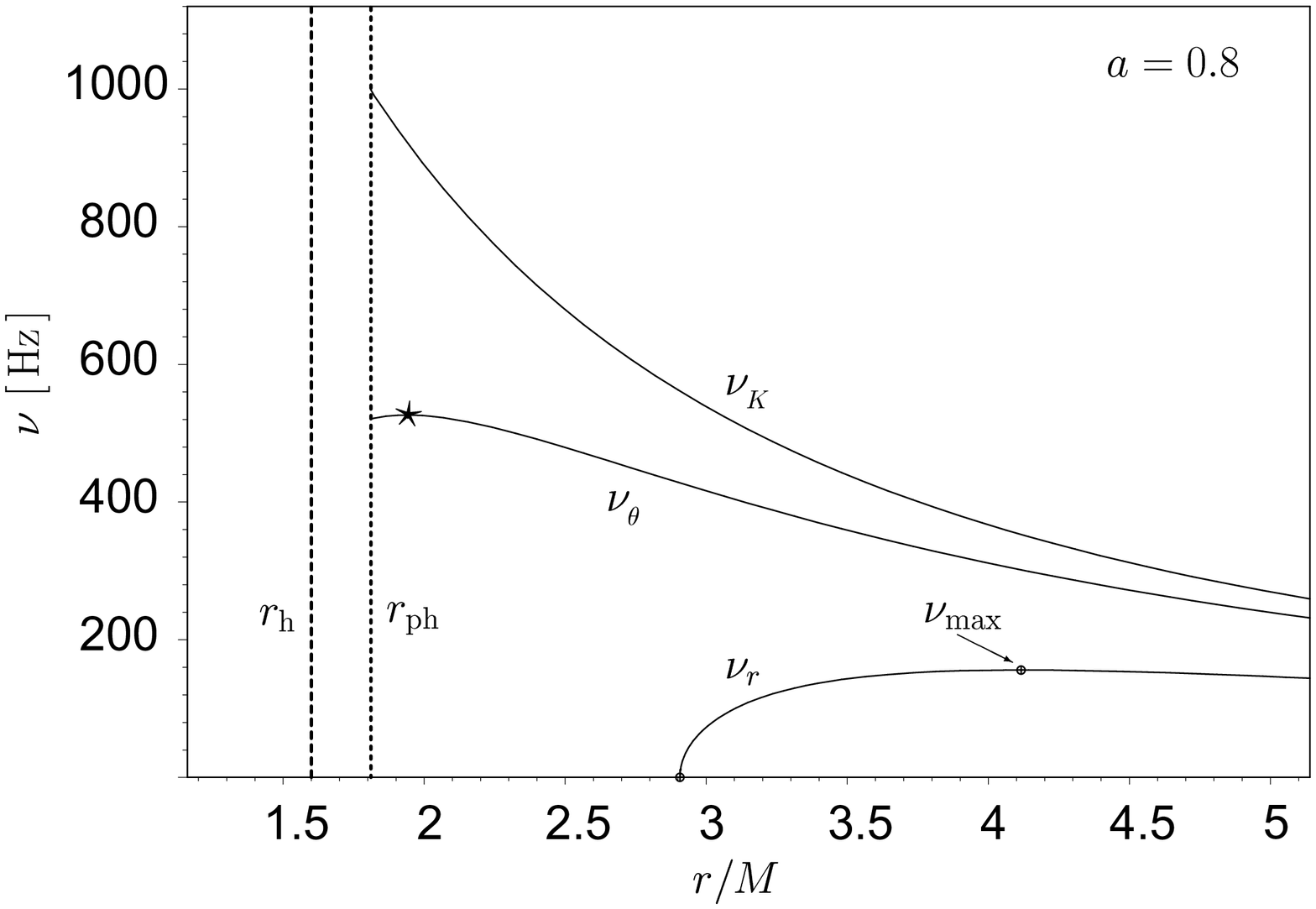}
  \caption{The same as in Figure 2, but for a $10\,M_{\odot}$ black hole
   with a moderately high spin, $a=0.8$.}
\end{figure}

For example, note that for black holes,
all three orbital frequencies: Keplerian $\nu_{\rm K}$, 
radial epicyclic $\nu_{r}$, and vertical epicyclic $\nu_{\theta}$, 
also have the general form

\begin{equation}
\nu = f (x, a)\left ( {{GM_0}\over {r_G^{~3}}}\right )^{1/2},
\end{equation}

\noindent with $a$ the dimensionless angular momentum of the black hole,
and $f(x, a)$ a dimensionless function, different for each frequency.
For all {\it relativistic} frequencies, $x = x(a)$ is fixed, 
and then the above formula predicts that frequencies scale
as $\nu = (1/M)F(a)$. 
In particular, each orbital resonance $n:m$ discussed later in
these contribution occurs at its own resonance radius $x_{n:m}(a)$, 
as shown in Figure 13, while at the marginally stable orbit
(ISCO) $\nu_r(r_{\rm ms})=0$.  All of the models discussed below
follow the $1/M$ scaling. 

\section{Non-resonant models}

\subsection{The highest possible orbital frequency}
\noindent
In an accretion disk, matter moves, roughly,  on circular orbits in
the region $r > r_{in}$  and free-falls in the region $r < r_{in}$.
The transition radius $r = r_{in}$, closely coinciding with the sonic
point, is often called the inner radius of the accretion disk.  Thin,
standard Shakura Sunyaev disks with their high efficiencies have an
inner edge located almost exactly at ISCO. In general, the inner edge
is located between the ISCO and the marginally bound  (RISCO, at
$r_{\rm mb}$) circular orbits \cite{26},  depending on the disk
efficiency. ADAFs, with their very low efficiencies,  have the inner
edges almost exactly at RISCO \cite{28}.  The same is true for
super-Eddington slim \cite{29} and thick \cite{26} disks. For a non-rotating
black hole one has $r_{\rm mb} = 4M$, ~$\nu_K(r_{\rm mb}) =
4037\,(M/M_{\odot})^{-1}\,$[Hz], ~and $r_{\rm ms} = 6M$, 
~$\nu_K(r_{\rm ms}) = 2197\,(M/M_{\odot})^{-1}\,$[Hz].


\begin{figure*}[!ht]
\includegraphics[angle=-90, width=78mm]{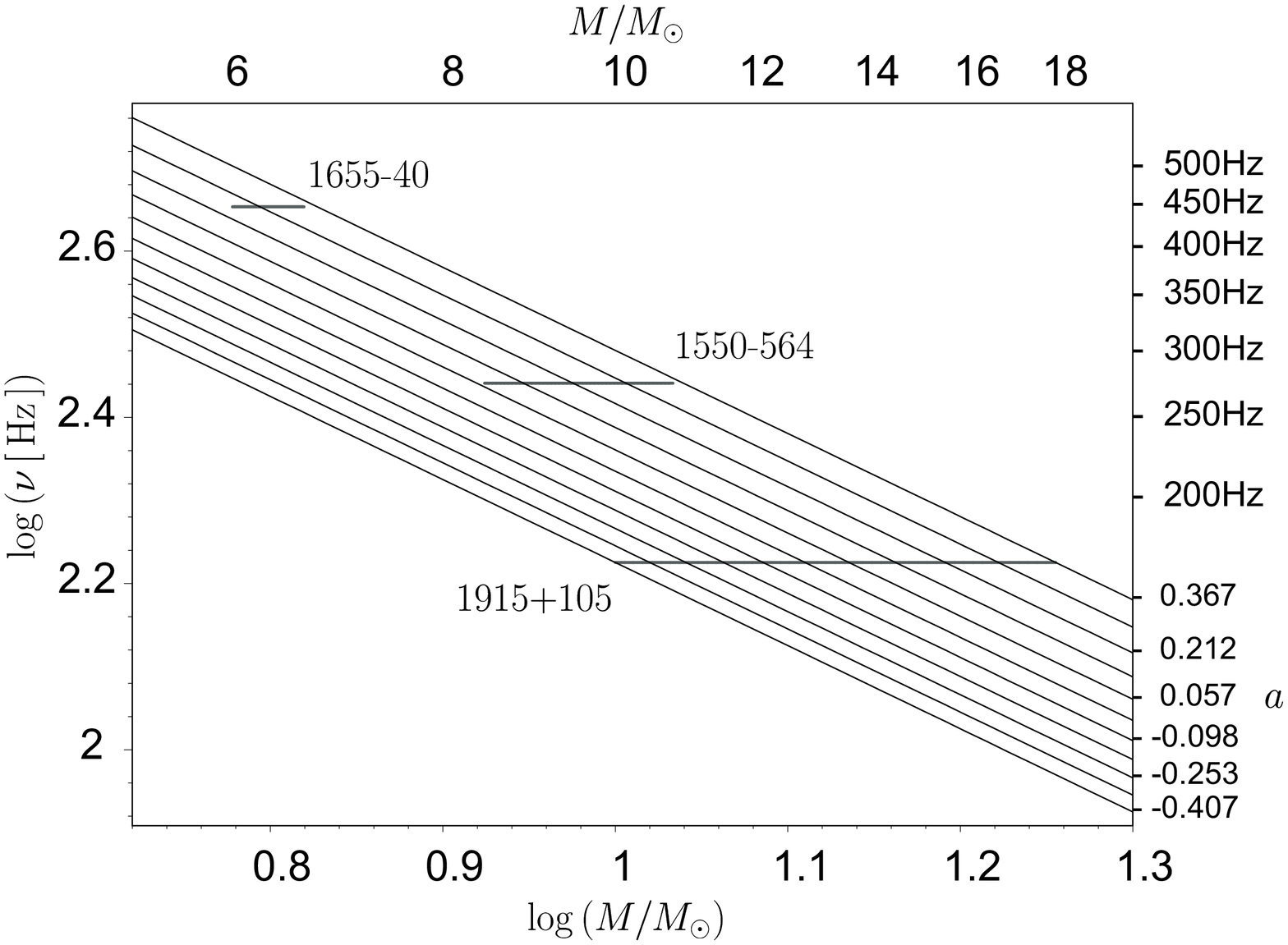}
\caption{ISCO orbital frequencies compared with the observed upper 
frequency of the hHz pair of QPOs.}
\end{figure*}


\begin{figure*}[!ht]
\includegraphics[angle=-90, width=78mm]{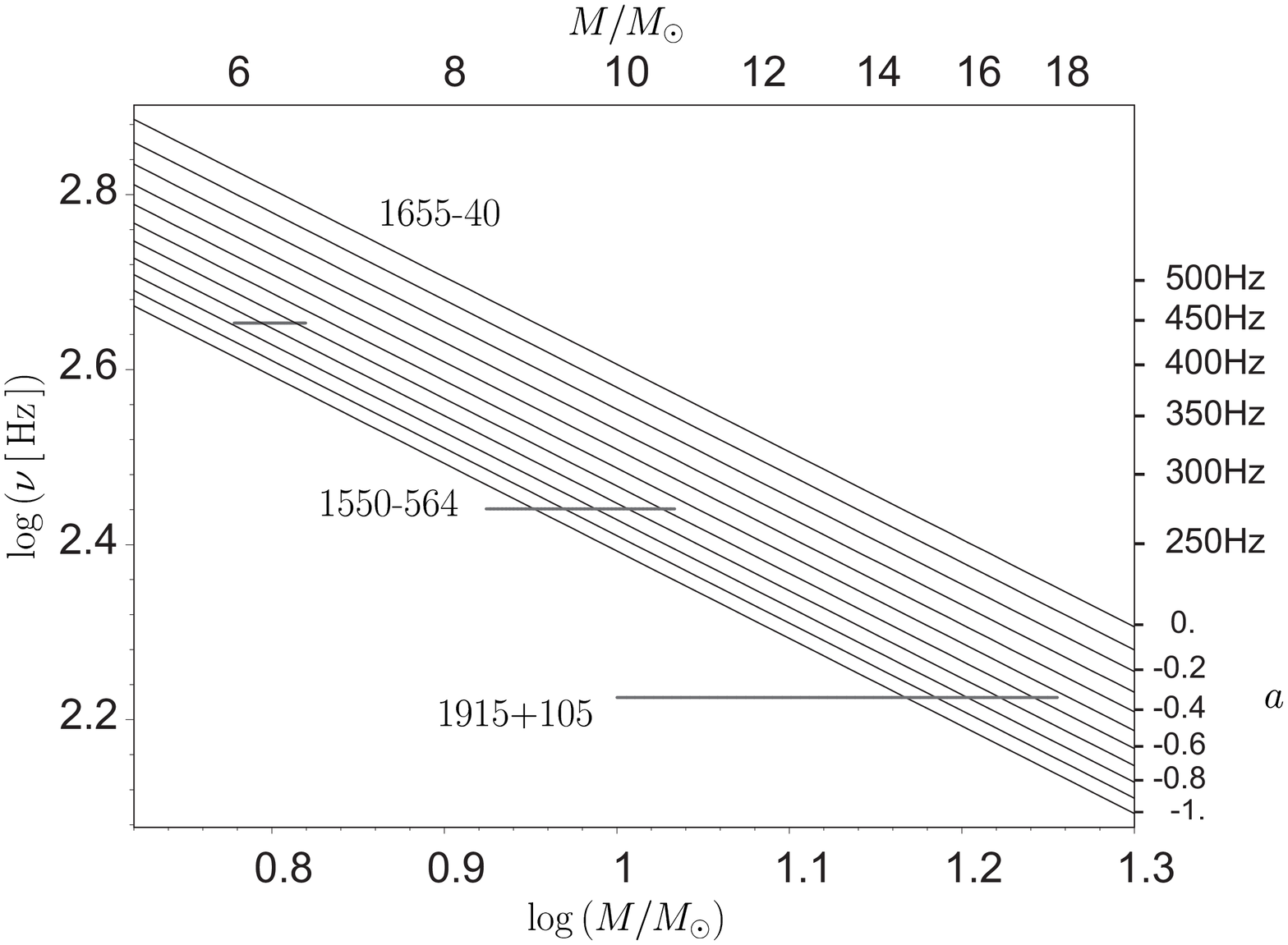}
\caption{RISCO orbital frequencies compared with the observed upper
frequency of the hHz pair of QPOs.}
\end{figure*}

 Let us ignore for the moment that the frequency pairs are in a
3:2 ratio.
Numerous studies have tried to infer the properties of neutron stars
on the assumption that the upper kHz QPO frequency is close to the ISCO
frequency (e.g., \cite{95}).
Identifying RISCO
frequencies with the upper frequency of the hHz pairs in microquasars,
whose mass is known, would require counter-rotating disks
(Fig.~5). Comparing the same frequency with the one at ISCO
leads to sensible but small values of the black hole spin (Fig.~4).
In most  models,
the QPO frequency corresponds to a characteristic frequency not very close
to the inner edge of the disk. This leads to substantially higher values of
the black hole rotation rate.

\subsection{The trapped modes}

One of the characteristic properties of the oscillations of
relativistic disks is the presence of trapped mode oscillations
\cite{16,18,31,35,ws,32,33}. The physical reason for the trapping is that
the radial epicyclic frequency, $\nu_r$, is not monotonic but has a
maximum value, $\nu_{\rm trapp}$, at a radius $r_{\rm trap}$
larger than the ISCO. For the non-rotating black hole hole
$r_{\rm trap} = 8 M$ ~\cite{18}. The g-mode (inertial-gravity)
oscillations \cite{32} can be characterized by a restoring force that is
typically dominated by the net gravitational-centrifugal force. The
axisymmetric ($m = 0$) g-modes are centered at $r_{\rm
trap}$. Non-axisymmetric trapped g-modes with the azimuthal
wave-number $m=1$ have frequencies \cite{35},


\begin{equation}
\nu \sim \nu_{\rm K} (r_{\rm trap}) \pm \nu_{\rm trapp}, 
~~~{\rm and} ~~~\nu \sim \nu_{\rm K} (r_{\rm trap}).
\end{equation}

\noindent In Figure 6 we show the highest frequency connected to these 
oscillations, $\nu_{\rm upp} = \nu_{\rm K} (r_{\rm trap}) + \nu_{\rm trapp}$,
 and compare it with observations.


\begin{figure*}[!ht]
\includegraphics[angle=-90, width=78mm]{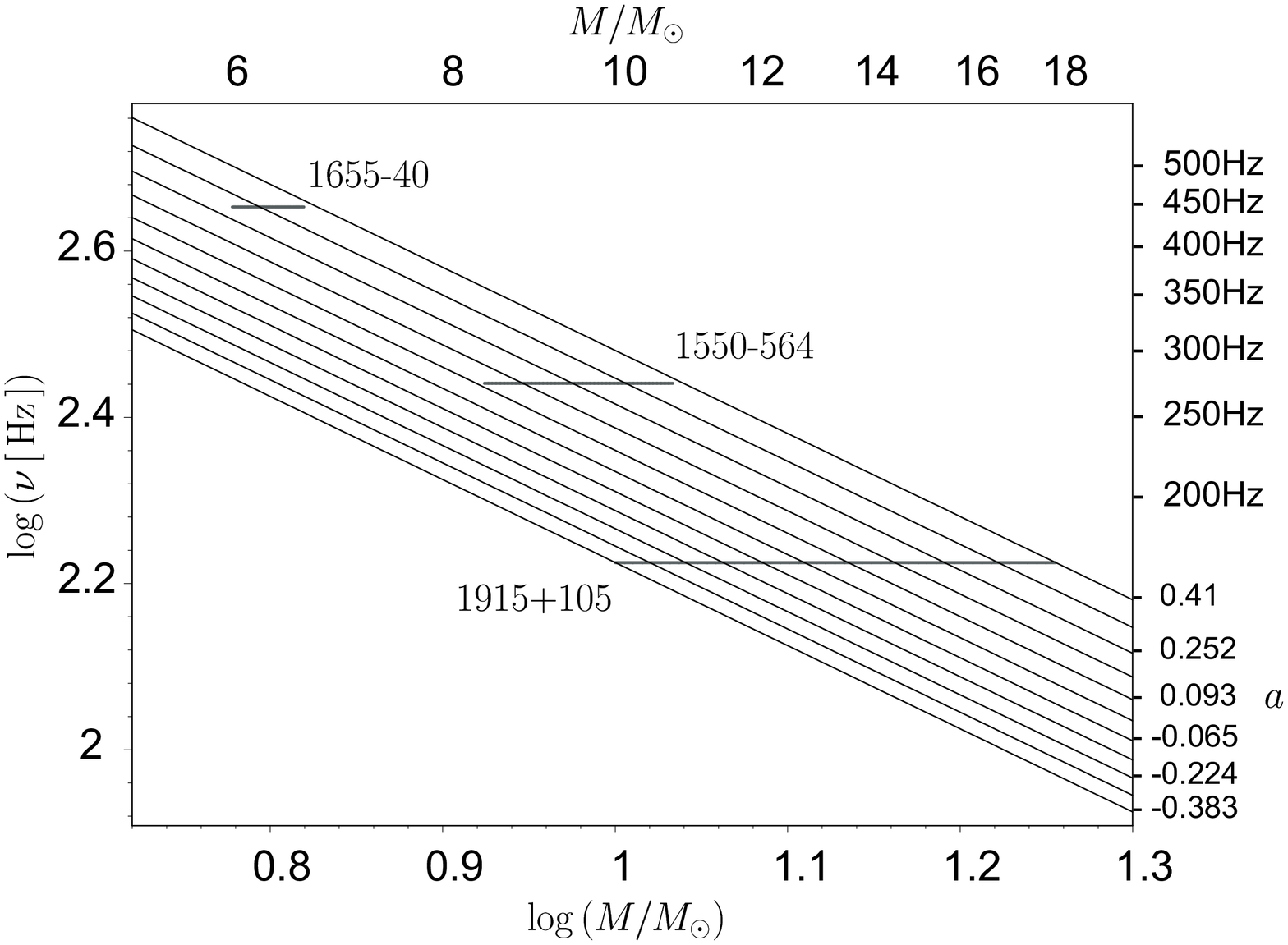}
\caption{Maximum frequency of the trapped, $m=1$, ~$g-$mode
compared with the upper QPO frequency.}
\end{figure*}


\begin{figure*}[!ht]
\includegraphics[angle=-90, width=78mm]{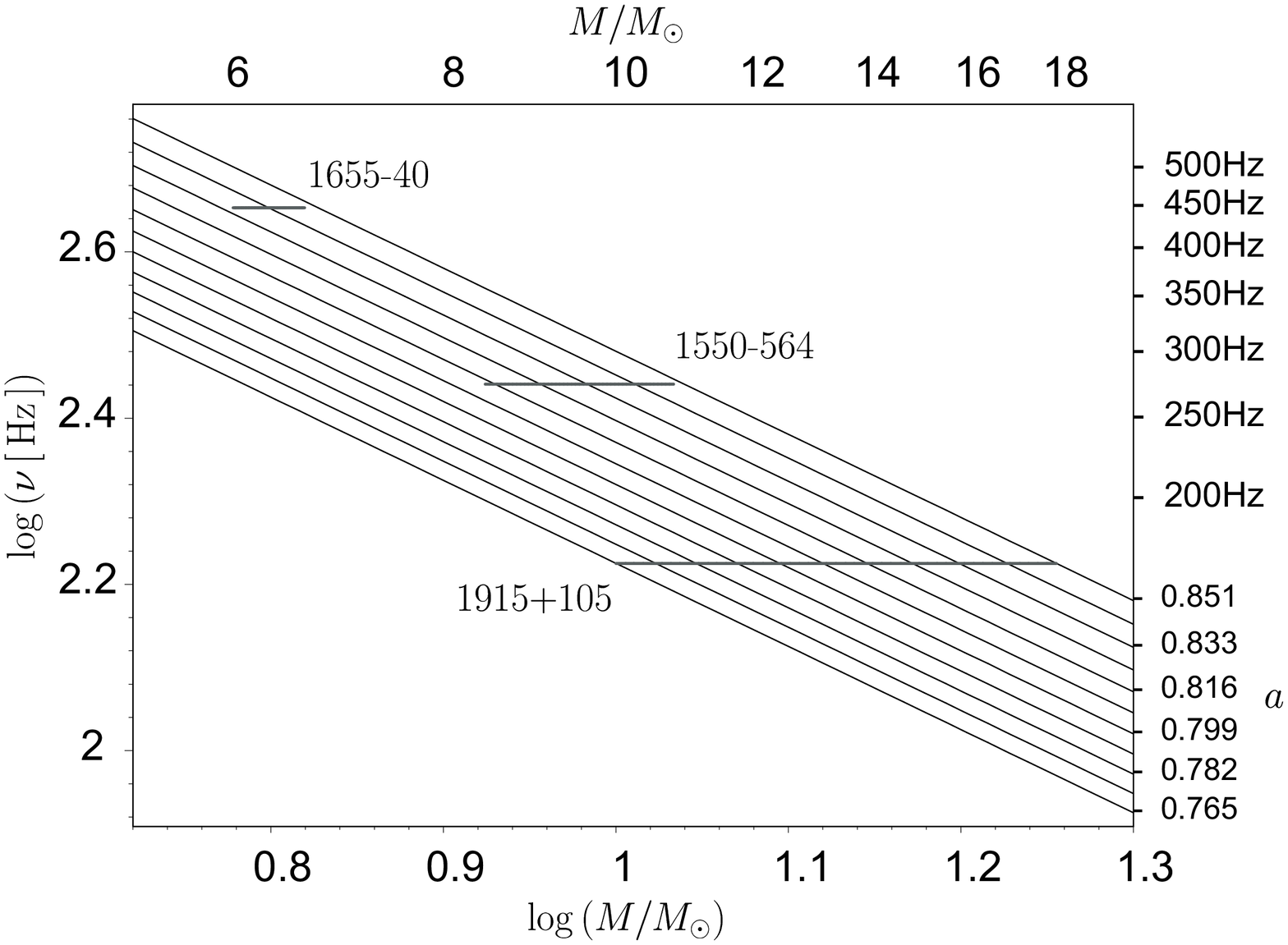}
\caption{Frequency of the c-mode at ISCO compared with the upper  QPO
frequency.}
\end{figure*}

\subsection{Dragging of inertial frames and the c-mode}

The `corrugation' c-mode \cite{33,31,30} is a non-axisymmetric, vertically
incompressible wave near the inner edge of the disk that exists only
for co-rotating disks, $a > 0$. It precesses around the angular
momentum of the black hole. Its frequency coincides with the
Lense-Thirring frequency produced by the dragging of inertial
frames. In Figure~7 we compare with observations the
highest frequency connected to the c-mode, assuming that the mode
locates at ISCO:

\begin{equation}
\nu_{\rm upp}= \nu_{\rm LT}(r_{\rm ms}) = \frac{ac}{\pi r_G}
(\frac{r_G}{r_{\rm ms}})^3.
\end{equation}
\noindent In reality, the mode is trapped further out in the disk, and
correspondingly, the inferred value of black hole spin is higher \cite{30}.
\section{}

\section{Non-linear, relativistic orbital resonances}

\subsection {``Keplerian'' resonances}

\noindent 
It is possible for the radial epicyclic frequency to be in a resonant
relation with the orbital frequency, $n\nu_r=m\nu_K$, with $n$, $m$ integer
\cite{9}.
For example, g-modes have pattern frequency
$2\pi\nu_m=\pm(\nu_r\pm m\nu_K)$,  and these can be in co-rotation resonance,
i.e., with $\nu_m=\nu_K$ \cite{96}. The case
$\nu_K/\nu_r =3/2$ is excluded by observations \cite{9} (Fig.~8).
The remaining possibilities \cite{9}
are that the upper frequency is $\nu_K\pm\nu_r$, with $\nu_K/\nu_r =2$,
or $\nu_K/\nu_r =3$, and $\nu_{\rm upper}/\nu_r =3/2$ (Figs.~9,~10).
However, co-rotation resonance leads to damping, and not excitation,
of modes \cite{97}.

Another possibility is based on the following idea \cite{23}. When the
potential vorticity is conserved, coherent vortices tend to form in
pairs with opposite vortices. One may imagine that because the spatial
distance between the two structures, which oscillates with the epicyclic
radial frequency, depends on the velocity profile of the disk,
i.e., also on the oscillations of orbital velocity, a resonance between
these two frequencies is possible.


\begin{figure*}[!ht]
\includegraphics[angle=-90, width=78mm]{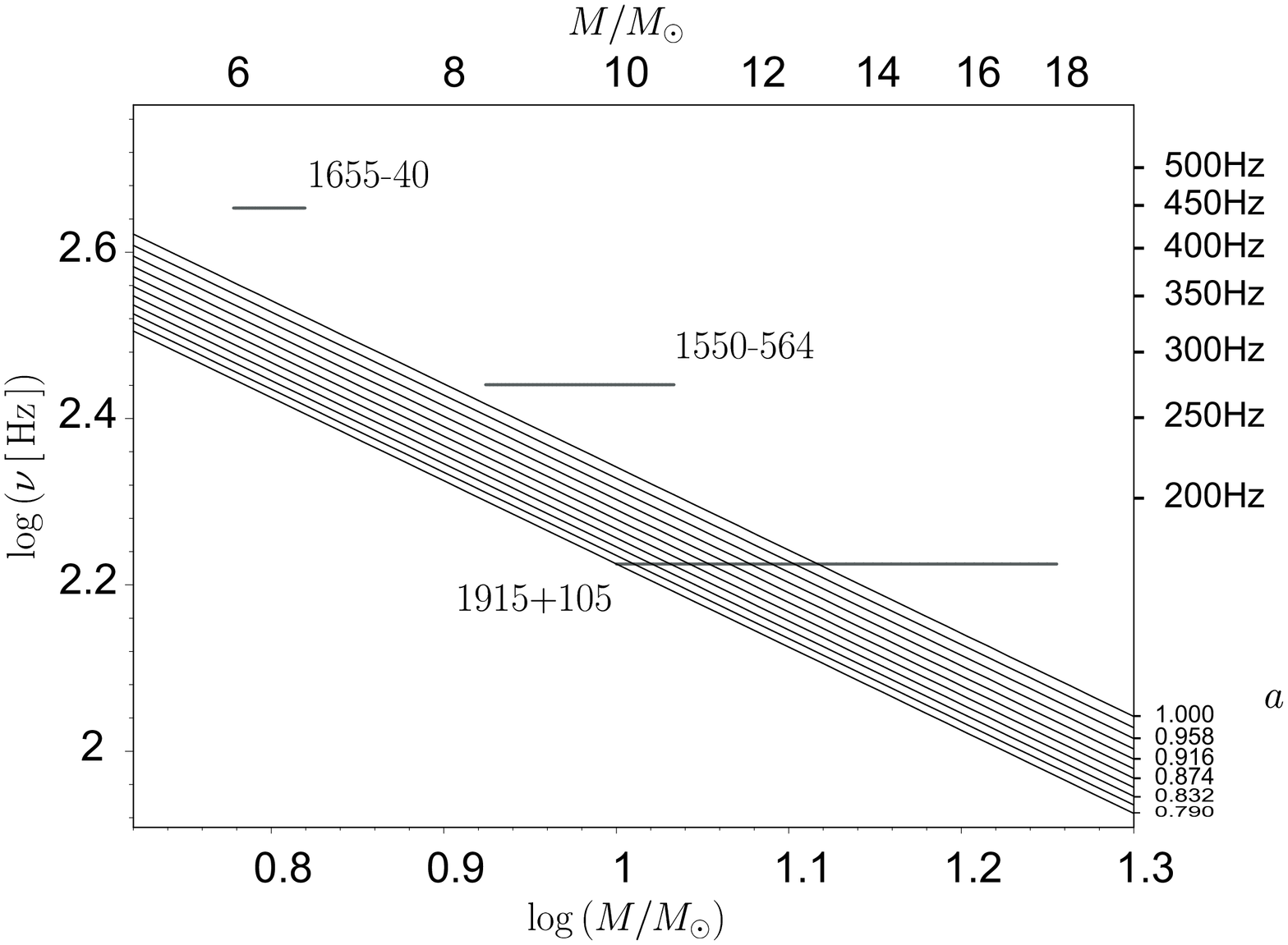}
\caption{Unsuccessful fit of observations to the Keplerian 3:2 resonance.}
\end{figure*}

\begin{figure*}[!ht]
\includegraphics[angle=-90, width=78mm]{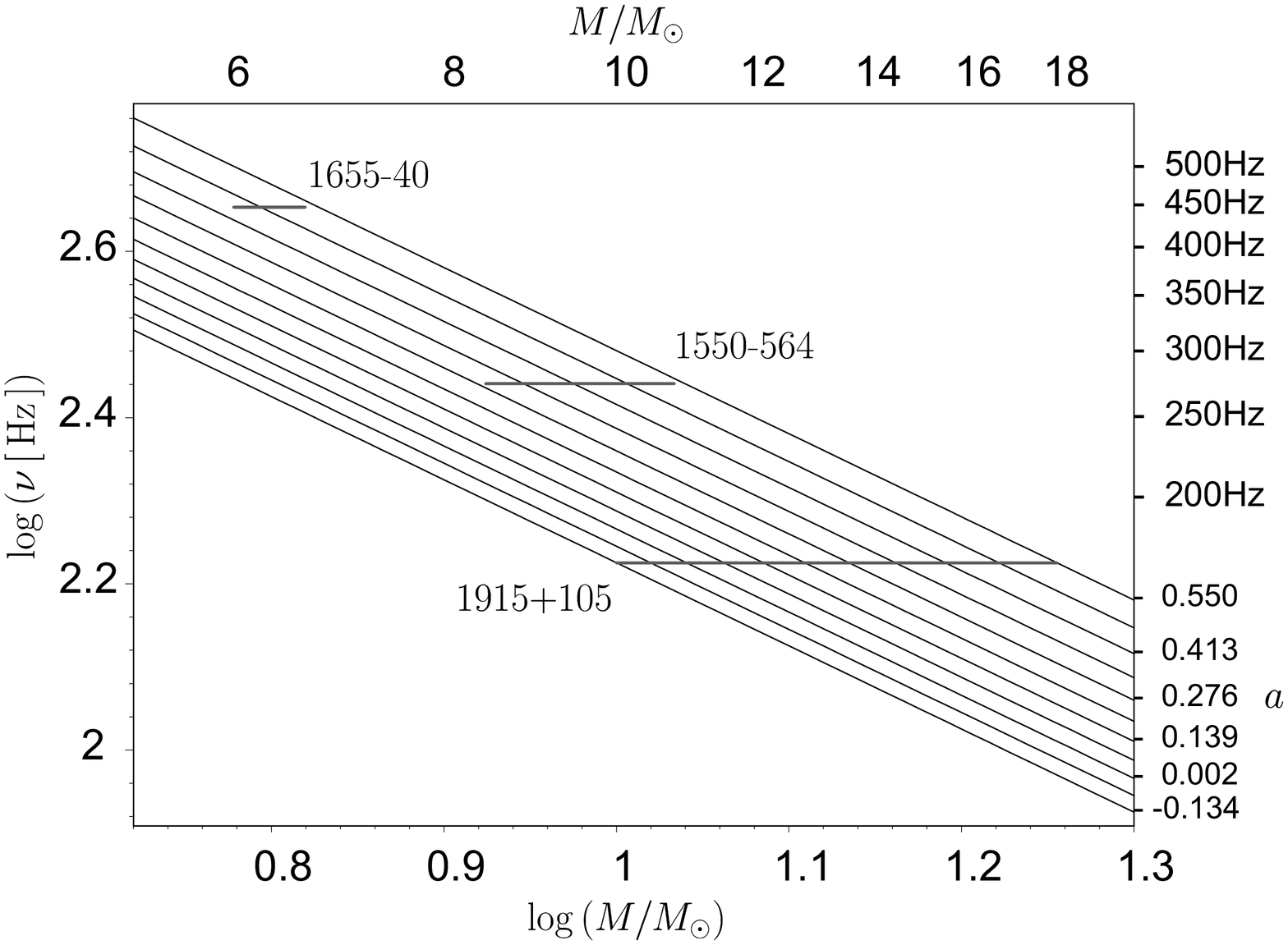}
\caption{Keplerian 3:1 resonance.}
\end{figure*}

\begin{figure*}[!ht]
\includegraphics[angle=-90, width=78mm]{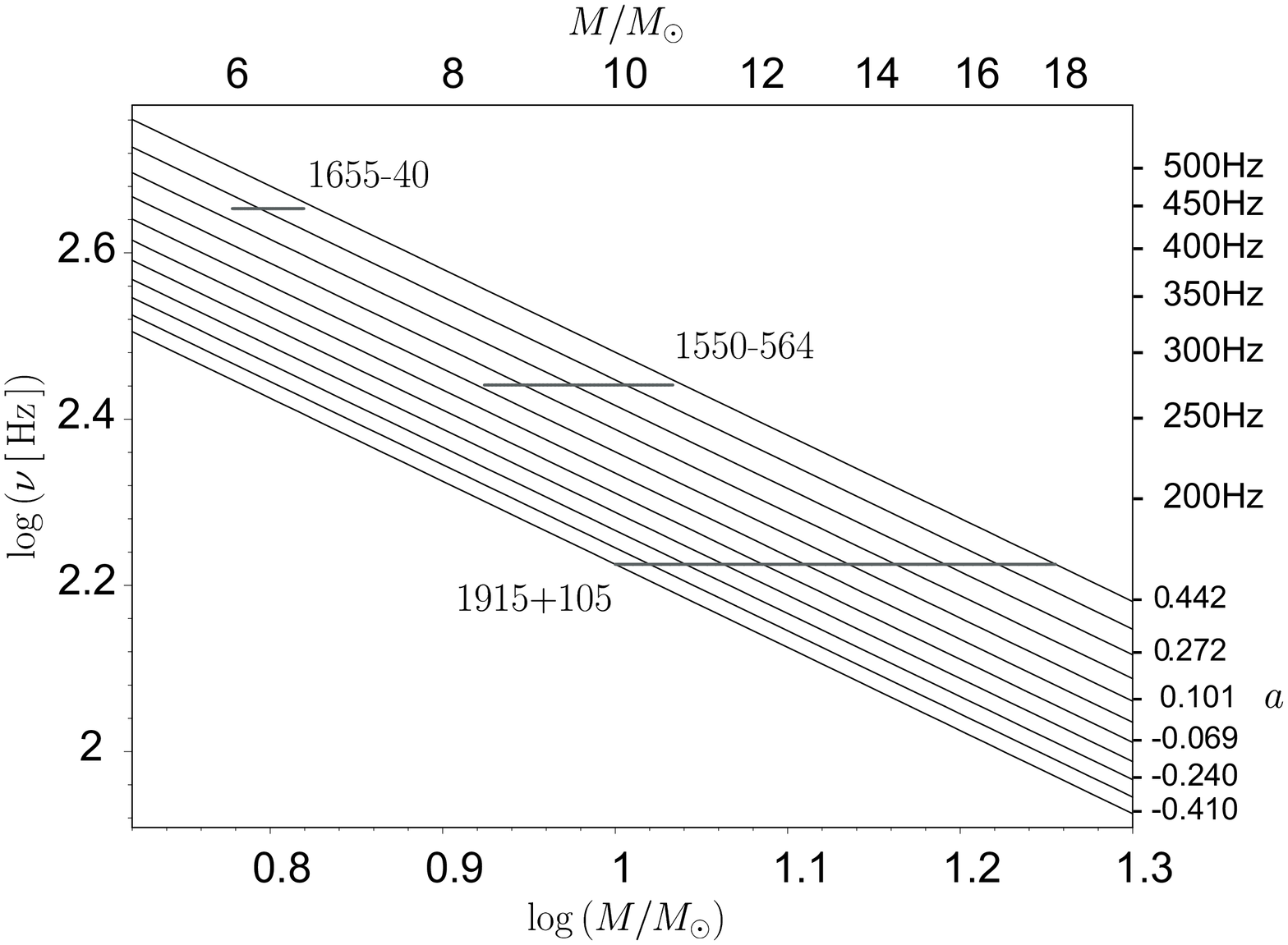}
\caption{Keplerian 2:1 resonance.}
\end{figure*}

\subsection{Epicyclic motions}
\noindent
The effective potential ${U}(r, \theta; \ell)$ for orbital motion of a
particle with a fixed angular momentum $\ell > \ell_{ms}$ has a
minimum at $r_0(\ell)$, corresponding to the location of a stable
circular orbit. The second order term in its Talyor expansion (for
simplicity we write it on the equatorial plane $\theta =\pi/2$)
gives the epicyclic frequencies, terms in the next order


\begin{equation}
{1\over2} \left ( {{\partial^2 {U}}\over {\partial r^2}}\right )_0
 (r - r_0)^2 + {1\over6} 
\left ( {{\partial^3 {U}}\over {\partial r^3}}\right )_0 (r - r_0)^3 + ...
\end{equation}

\noindent contain higher than quadratic terms, 
which means that small oscillations around the minimum at $r-r_0$ 
are described by non-linear differential equations \cite{19,20}. 
Non-linear resonances that may be excited in these non-linear oscillations 
have several characteristic properties that closely resemble those 
observed in QPOs.

 We consider two possibilities.

\subsection{The forced 3:1 and 2:1 resonances}

A direct resonant forcing of vertical oscillations by the radial ones
through a pressure coupling, and with $\delta r \sim \cos
(\omega_r\,t)$, is evident in recent numerical simulations of
oscillations of a perfect fluid torus \cite{13}. This supports
a possible model for the twin peak kHz QPOs: a forced
non-linear oscillator,


\begin{equation}
\delta \ddot \theta + \omega_{\theta}^2\delta \theta + O^2(\delta \theta ) =
 h \cos (\omega_r\,t), ~~~\omega_{\theta} \approx n\,\omega_r.
\end{equation}

\noindent Obviously, there is no integer value of $n$ such 
that $\omega_{\theta}$ and $\omega_r$ could be in the 3:2 ratio. 
However,  non-linear terms alow the presence of 
{\it combination frequencies} \cite{19, 20},


\begin{equation}
\omega_- = \omega_{\theta} - \omega_r, ~~~\omega_+ = \omega_{\theta} + \omega_r.
\end{equation}

\noindent One of these combination frequencies may be in a 3:2 ratio with
$\omega_{\theta}$ if and only if $n = 2$, or $n = 3$ in this forced resonance.
Simple arithmetic shows that in these two cases the observed frequencies
 $\nu_{\rm lower} = \omega_{\rm lower}/2\,\pi$ 
and $\nu_{\rm upper} = \omega_{\rm upper}/2\,\pi$ are uniquely given by,


\begin{equation}
[3:1]~~\omega_{\rm lower} =\omega_- = 2\,\omega_r, ~~~\omega_{\rm upper}
 = \omega_{\theta} = 3\,\omega_r,   
\end{equation}


\begin{equation}
[2:1]~~\omega_{\rm upper} = \omega_+ = 3\,\omega_r, ~~~\omega_{\rm lower} 
=  \omega_{\theta} = \,2\omega_r.
\end{equation}

\noindent We fit observed QPOs to these predicted by the forced
 epicyclic 3:1 and 2:1 resonances in Figures 11 and 12.


\begin{figure*}[ht]
\includegraphics[angle=-90, width=78mm]{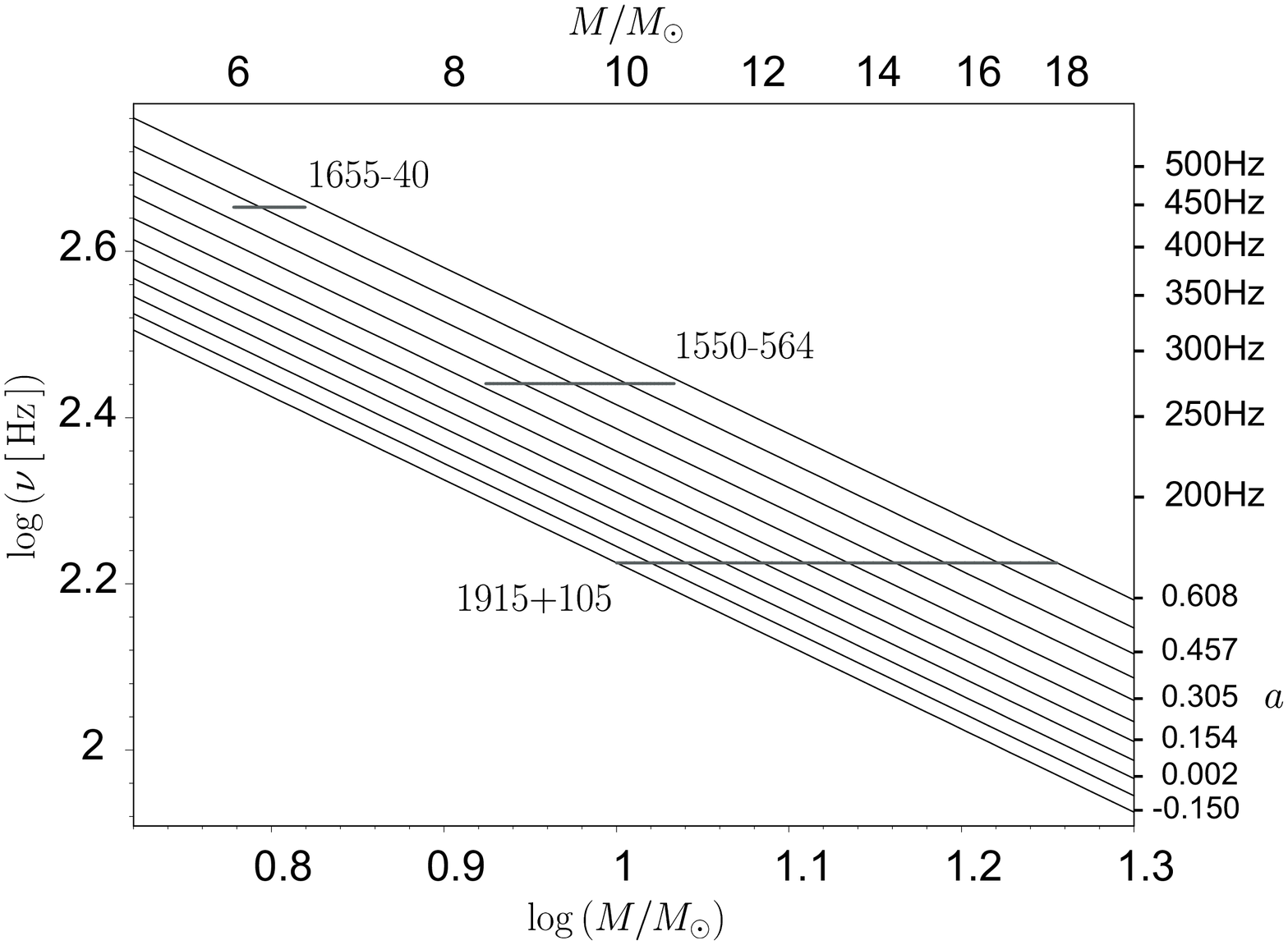}
\caption{Fit of the predictions of the 3:1 epicyclic forced resonance model to
observations.}
\end{figure*}


\begin{figure*}[ht]
\includegraphics[angle=-90, width=78mm]{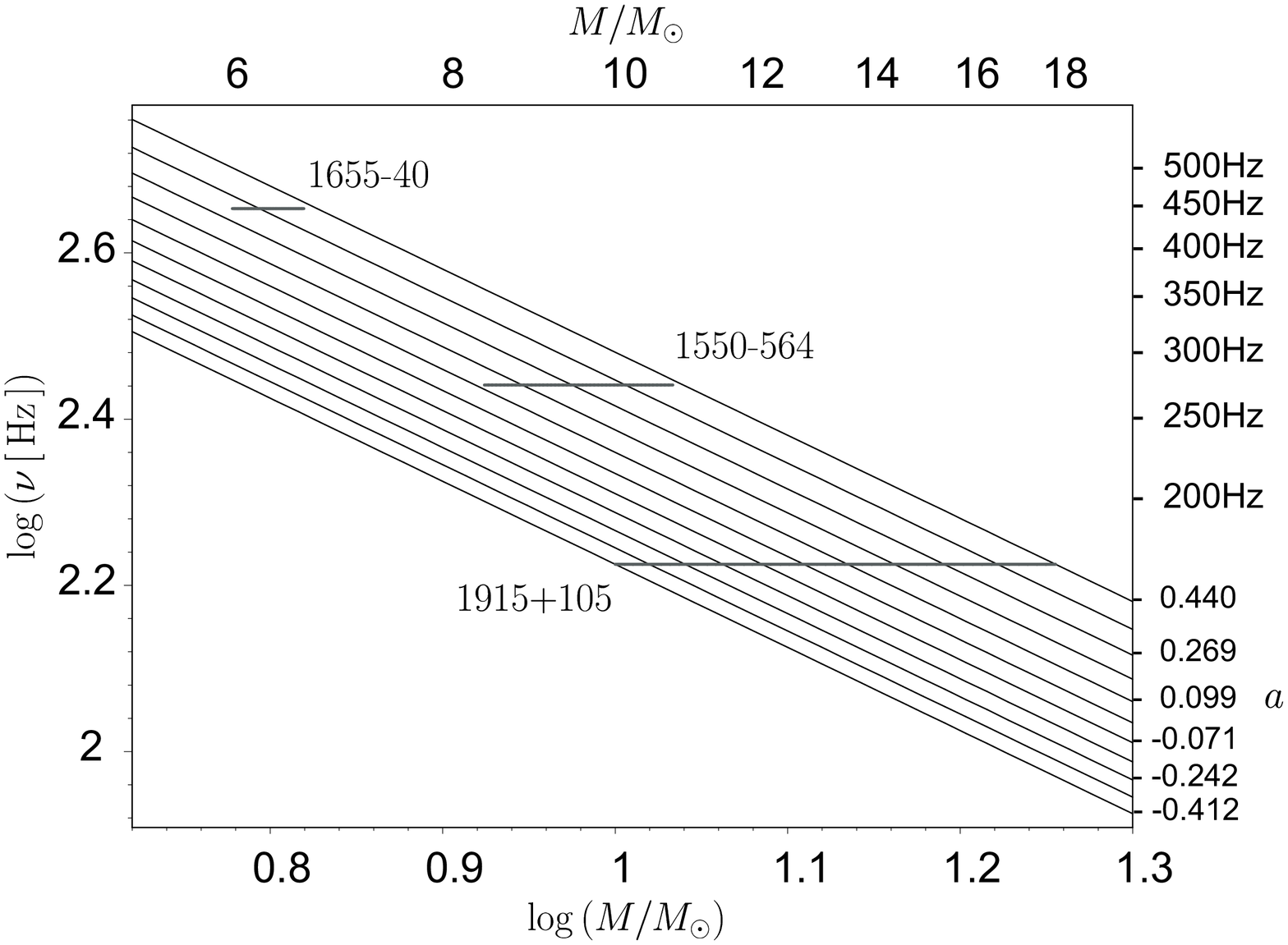}
\caption{The 2:1 forced epicyclic resonance.}
\end{figure*}

\subsection{3:2 parametric resonance}
\begin{figure*}[!ht]
\includegraphics[angle=-90, width=78mm]{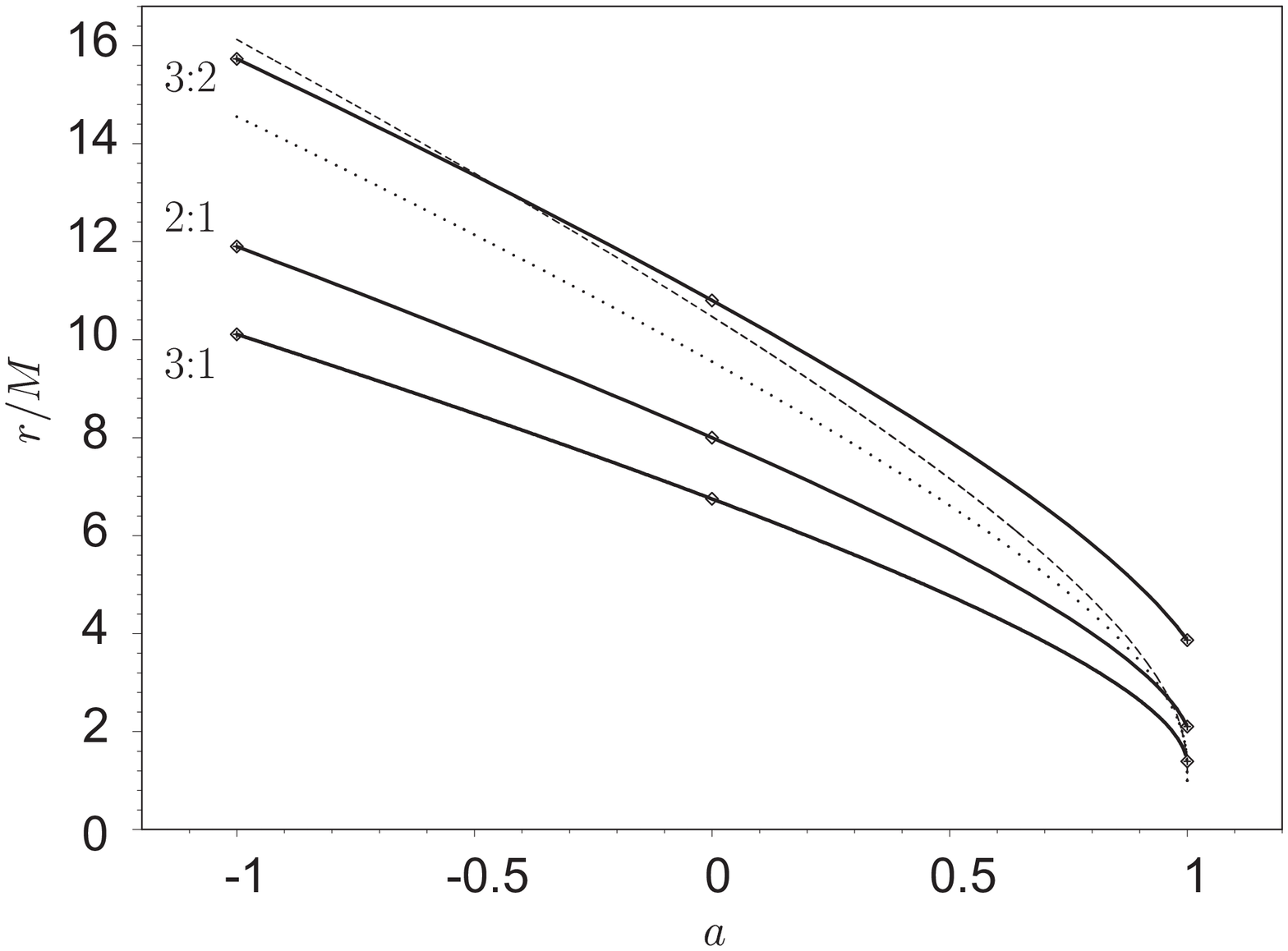}
\caption{The radii of three epicyclic resonances, 3:2, 3:1 and
2:1. Also shown are the location of the centre of
largest possible accretion torus (dashed lines), and the
location  (dotted lines), closer to the horizon, 
of the maximal locally emitted flux from the standard thin disk.}
\end{figure*}

\noindent We shall start with an argument appealing to physical
intuition and showing that the resonance to be discussed now is a very
natural, indeed necessary, consequence of strong gravity. In thin
disks, random fluctuations have $\delta r \gg \delta \theta$. Thus,
$\delta r \delta \theta$ is a first order term in $\delta \theta$ and
should be included in the first order equation for vertical
oscillations (\ref{eq_mot_vdir}). The equation now takes the form,


\begin{equation}
\label{eq_delta_a_theta}
\delta \ddot \theta +
 \omega_{\theta}^2\left [ 1 + h\,\delta r \right ] \delta \theta = 
\delta a_{\theta},
\end{equation}

\noindent where $h$ is a known constant. The first order equation for
$\delta r$ has the solution $\delta r = A_0 \cos (\omega_r\,t)$. 
Inserting this in (\ref{eq_delta_a_theta}) together with
$\delta a_{\theta} = 0$, one arrives at the Mathieu equation ($A_0$ is
absorbed in $h$),


\begin{equation}
\delta \ddot \theta + 
\omega_{\theta}^2\left [ 1 + h \,\cos (\omega_r \,t)\right ] \delta \theta =
 0, 
\end{equation}

\noindent that describes the {\it parametric resonance}. 
From the theory of the Mathieu equation one knows that when


\begin{equation}
{\omega_r \over \omega_{\theta}} = {\nu_r \over \nu_{\theta}} \approx
 {2 \over n}, ~~~~n =1,\,2, \,3 ..., 
\end{equation}

\noindent the parametric  resonance is excited \cite{19, 20}. The
resonance is strongest for the smallest possible value of
$n$. Because near black holes $\nu_r < \nu_{\theta}$, the smallest
possible value for resonance is $~n = 3$, which means that
$2\,\nu_{\theta} = 3\,\nu_r$. This explains \cite{99} the observed 3:2 ratio,
if, obviously,

\begin{equation}
\nu_{\rm upper} = \nu_{\theta},~~~~\nu_{\rm lower} = \nu_r. 
\end{equation}

\noindent Parametric resonance of
the type discussed above was found in numerical simulations of
oscillations in a nearly Keplerian accretion disk by Abramowicz et
al. \cite{11} and confirmed by exact analytic solutions \cite{21,22}. 
The analytic solution is accurate up to third order terms in
$\delta r$, $\delta \theta$, and based on the method of multiple
scales \cite{20}. Existence of the 3:2 parametric resonance is
therefore a mathematical poroperty of thin, nearly Keplerian disks. 


\begin{figure*}[!ht]
\centering
\includegraphics[angle=-90, width=78mm]{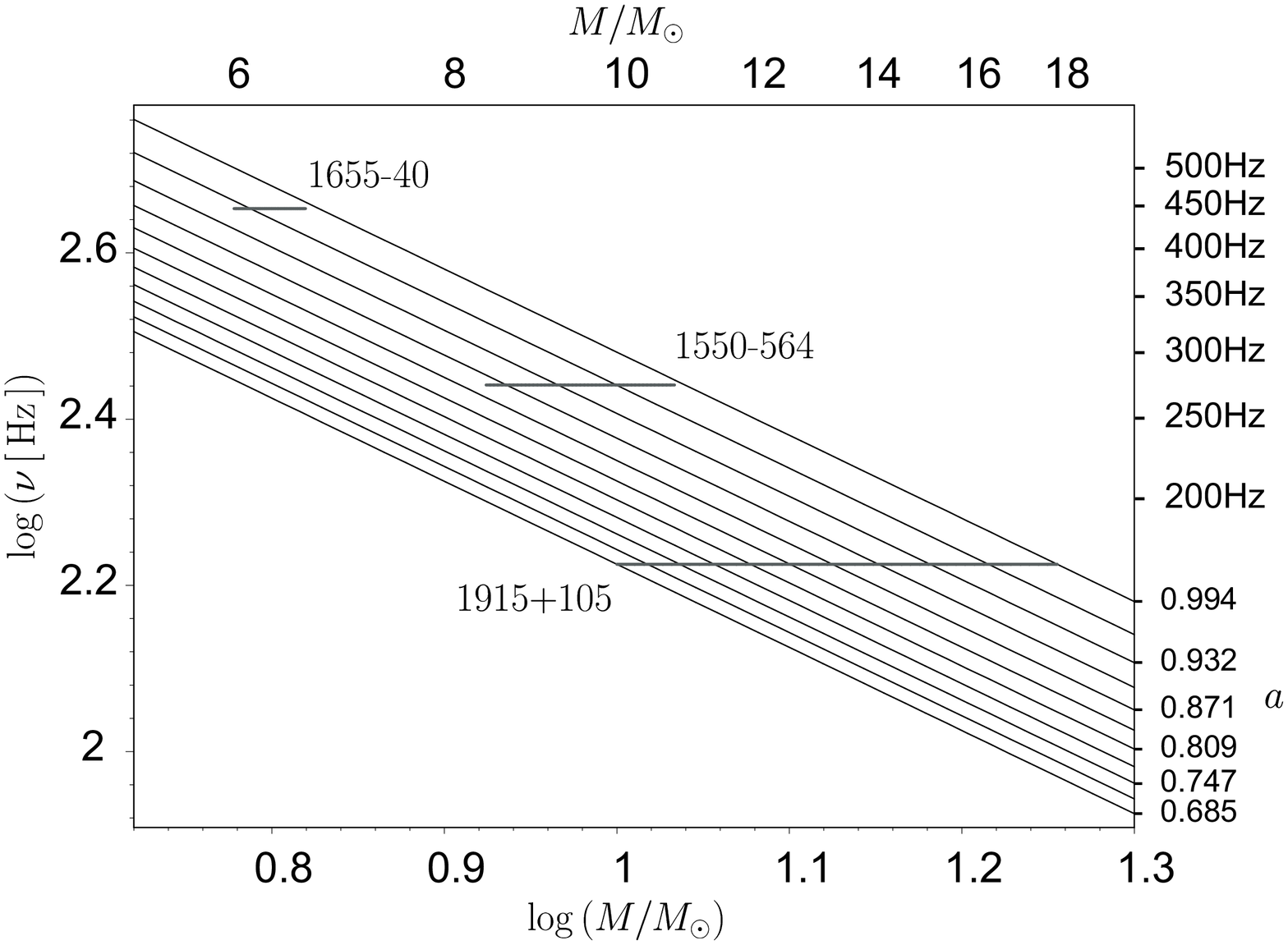}
\caption{\label{plot_3_2}Fit of the frequency 
$\nu_{upp} =\nu_{\theta} (M, a)$
 predicted by the 3:2 parametric resonance model
to the frequencies observed in three microquasars with known
masses. The spin parameter $a$ is not known from observations, and the
lines $a =\,$ const are calculated from the model. One should notice
that the deduced black hole spins are rather high, and fall in a rather narrow
range.}
\end{figure*}

It
was found that the resonance is exited only in the non-Keplerian case,
with some weak forces $\delta a_{\theta} \not = 0$ and $\delta a_r
\not = 0$ present. Their origin is certainly connected to stresses
(pressure, magnetic field, viscosity), but how exact details remain to
be determined --- at present $\delta a_{\theta}$ and $\delta a_r$ are
not calculated from first principles but described by an
ansatz\footnote{While the lack of a full physical understanding is
obviously not satisfactory, the experience tells that such a situation
is not uncommon for non-linear systems. Examples are known of
mathematically possible resonances causing damage in bridges,
areoplane wings etc., for which no specific physical excitation
mechanism could have been pinned down \cite{20}.}.
Of course in real disks neither $\delta r = A_0 \cos
(\omega_r \,t)$, nor $\delta a_{\theta} = 0$ exactly, but one may
expect that because these equations are approximately obeyed for thin
disks, the parametric resonance will also be excited in realistic
situations. And this is indeed the case \cite{34}. 

The parametric resonance occurs at a particular radius
$r_{3:2}(a)$, determined by the condition 
$3\omega_r (r_{3:2}, a) = 2\omega_{\theta} (r_{3:2}, a)$.
 We show the function $r_{3:2}(a)$ in
Figure 13. In Figure \ref{plot_3_2} we fit
the 3:2 resonance theoretically predicted frequencies to the
observational data for the three microquasars with known masses.
The scatter for the particular 3:2 resonance is not very
large because (Fig.~13) this resonances
occurs at $x_{3:2}(a) > 4$, where the influence of $a$ is not
dominant. 


\begin{table} [!ht]
\begin{tabular}{llll}
\hline
    \tablehead{1}{r}{b}{Microquasar}
  & \tablehead{1}{r}{b}{3:2 }
  & \tablehead{1}{r}{b}{2:1 }
  & \tablehead{1}{r}{b}{3:1 } \\
  
\hline
~~XTE  1550-564           & ~~0.94 & ~~0.27 & ~~0.46 \\   

~~GRO  1655-40            & ~~0.96 & ~~0.36 & ~~0.55 \\ 

~~GRS  1915+105           & ~~0.84 & ~~0.02 & ~~0.23 \\

\hline
\end{tabular}
\caption{Black hole spin in three microquasars calculated by fitting 
observations to the three resonance models}
\label{table-1}
\end{table}

\begin{figure*}[!ht]
\includegraphics[angle=-90, width=78mm]{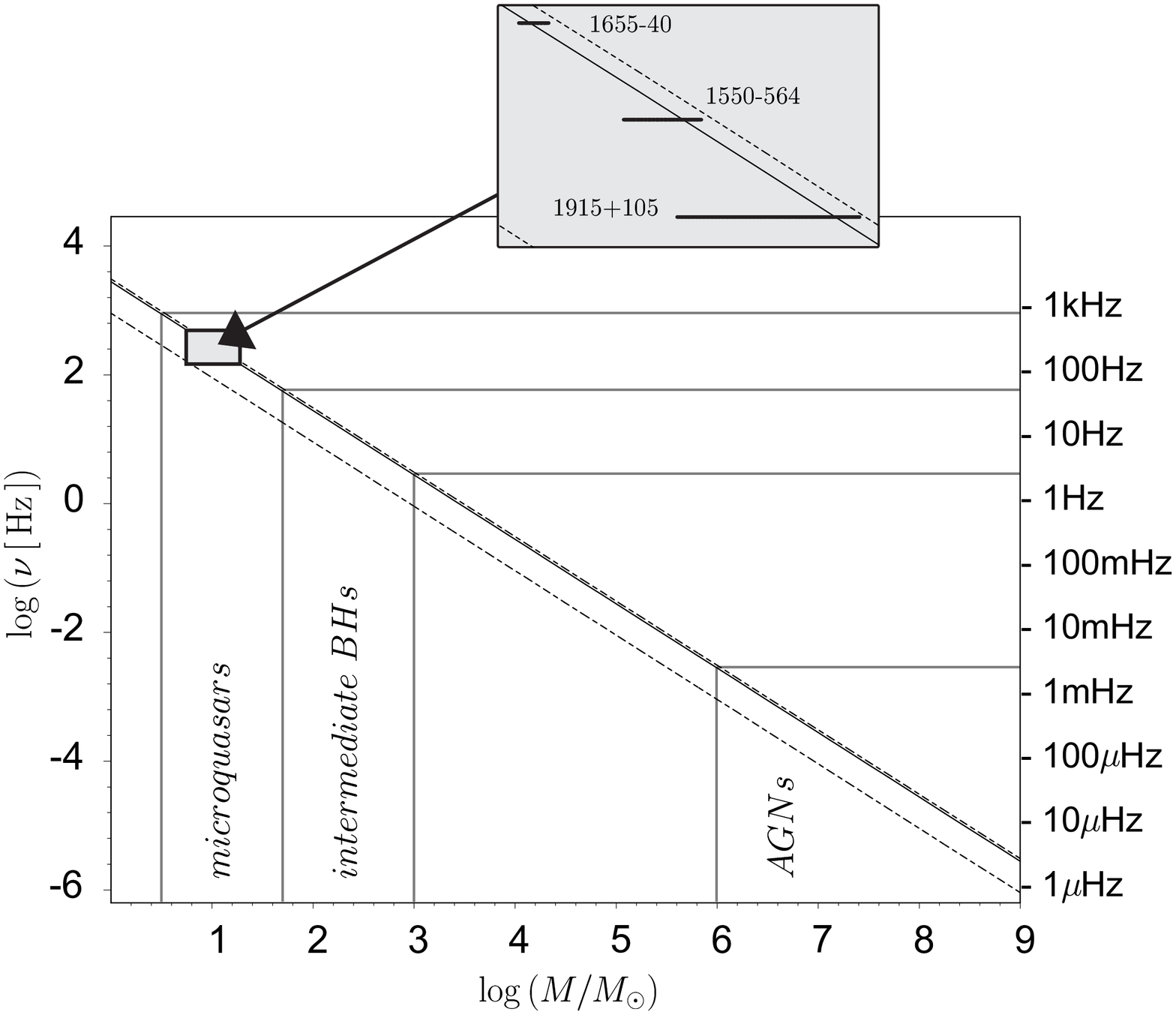}
\caption{The 1/M scaling of the frequencies of 3:2 QPO pairs
 may be used for mass estimates (from Abramowicz, Klu{\'z}niak, McClintock,
 \& Remilard, 2003).}
\end{figure*}

\section{Applications}
The $1/M$ scaling of the twin peak QPOs frequencies with the
3:2 ratio, was proposed by Abramowicz, Klu\'zniak, McClintock \&
Remillard \cite{98} as a method for estimating black hole masses in AGNs
and the recently discovered ultraluminous X-ray sources (ULXs),
 based on Mirabel's analogy between microquasars in our
Galaxy and distant quasars \cite{25}. Indeed, if the analogy is also valid
for accretion disk oscillations, then discovering in ULXs the twin
peak QPOs frequencies with the 3:2 ratio, would resolve the
controversy about their mass: if ULXs black holes have the same masses
as microquasars, the frequencies will be $\sim 100\,$Hz; if ULXs black
holes are $\sim 1000$ times more massive, the frequencies will be
$\sim 0.1\,$Hz instead.



\begin{theacknowledgments}
We thank Gabriel T\"or\"ok for preparing all the Figures and other
technical help. All Figures, except Figure 15, are taken from
Abramowicz, Klu\'zniak, Stuchlik, \& T\"or\"ok \cite{909}. Figure 15
is taken from our work with R.~Remillard \& J.~McClintock
\cite{98}. Most of the work reported here was done at the Silesian
University of Opava, in the Czech Republic, and at the Astrophysical
Fluids Facility in Leicester University, England. It was supported by
the European Commission grant {\it Access to Research Infrastructure
action of the Improving Human Potential Program} and by the Polish KBN
grant 2P03D01424. We thank  J.~Almergren, M.~Bursa, J.~Horak,
V.~Karas, F.~Lamb, J.-P.~Lasota, W.~Lee,  C.~Mauche, J.~McClintock,
R.~Remillard,  L.~Rezzolla,  J.~Schnittman,  E.~Spiegel,  P.~Rebusco,
M.~van~der~Klis, and R.~Wagoner, for their sugestions and comments on
this presentation.
\end{theacknowledgments}

\end{document}